\documentclass[aps,prl,superscriptaddress,floatfix,reprint]{revtex4-1}  

\usepackage{graphicx}
\usepackage{bm}
\usepackage{amssymb}
\usepackage{amsmath}
\usepackage{float}
\usepackage{pstricks} 
\usepackage{color}
\usepackage{bbold}
\usepackage{xr} 

\newcounter{subfigure}[figure]

\newcommand{\bra}[1]{\left\langle #1 \right\vert}
\newcommand{\ket}[1]{\left\vert #1 \right\rangle}

\newcommand{\grad}[1]{\nabla }

\graphicspath{{figures/}}

\begin{document}

\title{Quantum teleportation on a photonic chip}

\author{Benjamin~J.~Metcalf}
\email{b.metcalf1@physics.ox.ac.uk}
\affiliation{Clarendon Laboratory, University of Oxford, Parks Road, Oxford OX1 3PU, UK}

\author{Justin~B.~Spring}
\affiliation{Clarendon Laboratory, University of Oxford, Parks Road, Oxford OX1 3PU, UK}

\author{Peter~C.~Humphreys}
\affiliation{Clarendon Laboratory, University of Oxford, Parks Road, Oxford OX1 3PU, UK}

\author{Nicholas~Thomas-Peter}
\affiliation{Clarendon Laboratory, University of Oxford, Parks Road, Oxford OX1 3PU, UK}

\author{Marco~Barbieri}
\affiliation{Clarendon Laboratory, University of Oxford, Parks Road, Oxford OX1 3PU, UK}

\author{W.~Steven~Kolthammer}
\affiliation{Clarendon Laboratory, University of Oxford, Parks Road, Oxford OX1 3PU, UK}

\author{Xian-Min~Jin}
\affiliation{Clarendon Laboratory, University of Oxford, Parks Road, Oxford OX1 3PU, UK}
\affiliation{Department of Physics and Astronomy, Shanghai Jiao Tong University, Shanghai 200240, PR China}

\author{Nathan~K.~Langford}
\affiliation{Department of Physics, Royal Holloway, University of London, TW20 0EX, UK}

\author{Dmytro~Kundys}
\affiliation{Optoelectronics Research Centre, University of Southampton, Southampton, SO17 1BJ, UK}
\affiliation{School of Physics and Astronomy, University of Manchester, Oxford Road, Manchester, M13 9PL, UK}

\author{James~C.~Gates}
\affiliation{Optoelectronics Research Centre, University of Southampton, Southampton, SO17 1BJ, UK}

\author{Brian~J.~Smith}
\affiliation{Clarendon Laboratory, University of Oxford, Parks Road, Oxford OX1 3PU, UK}

\author{Peter~G.~R.~Smith}
\affiliation{Optoelectronics Research Centre, University of Southampton, Southampton, SO17 1BJ, UK}

\author{Ian~A.~Walmsley}
\affiliation{Clarendon Laboratory, University of Oxford, Parks Road, Oxford OX1 3PU, UK}

\date{\today}

\maketitle

{\bf
Quantum teleportation is a fundamental concept in quantum physics~\cite{Bennett1993} which now finds important applications at the heart of quantum technology including quantum relays~\cite{Jacobs2002,Collins2005}, quantum repeaters~\cite{Briegel1998} and linear optics quantum computing (LOQC)~\cite{Knill2001,Gottesman1999}. Photonic implementations have largely focussed on achieving long distance teleportation due to its suitability for decoherence-free communication~\cite{Ma2012,Jin2010,Marcikic2003}. Teleportation also plays a vital role in the scalability of photonic quantum computing~\cite{Knill2001,Gottesman1999}, for which large linear optical networks will likely require an integrated architecture. Here we report the first demonstration of quantum teleportation in which all key parts---entanglement preparation, Bell-state analysis and quantum state tomography---are performed on a reconfigurable integrated photonic chip. We also show that a novel element-wise characterisation method is critical to mitigate component errors, a key technique which will become increasingly important as integrated circuits reach higher complexities necessary for quantum enhanced operation.}

Quantum teleportation is essential to many schemes for universal fault-tolerant quantum computation, making it an important protocol for any physical implementation of a quantum information processor~\cite{Childs2005,Kok2007}. In their seminal work, Knill, Laflamme, and Milburn showed that such a quantum processor could be constructed using only linear optical elements, at the expense of rendering each quantum logic gate probabilistic~\cite{Knill2001}. Adapting the teleportation scheme of Gottesman and Chuang~\cite{Gottesman1999}, they then showed that this protocol could be efficiently scaled to a large number of concatenated gates, motivating a renewed interest in building more complex linear optical circuits for quantum information processing~\cite{Kok2007}. Realizing such a scheme requires building large, sophisticated networks of nested optical interferometers. This motivates the use of waveguides integrated onto compact and inherently stable photonic chips, and pioneering work has shown the viability of this approach for two-~\cite{Shadbolt2011,smith2009,Crespi2011} and three-photon interference experiments~\cite{Metcalf2013,Spring2012,Crespi2013}. These latter works highlighted the problems caused by photon loss, low data rates, and fabrication imperfections which make the extension to even higher photon numbers far from straightforward.

\begin{figure*}
 \includegraphics[width=1\textwidth]{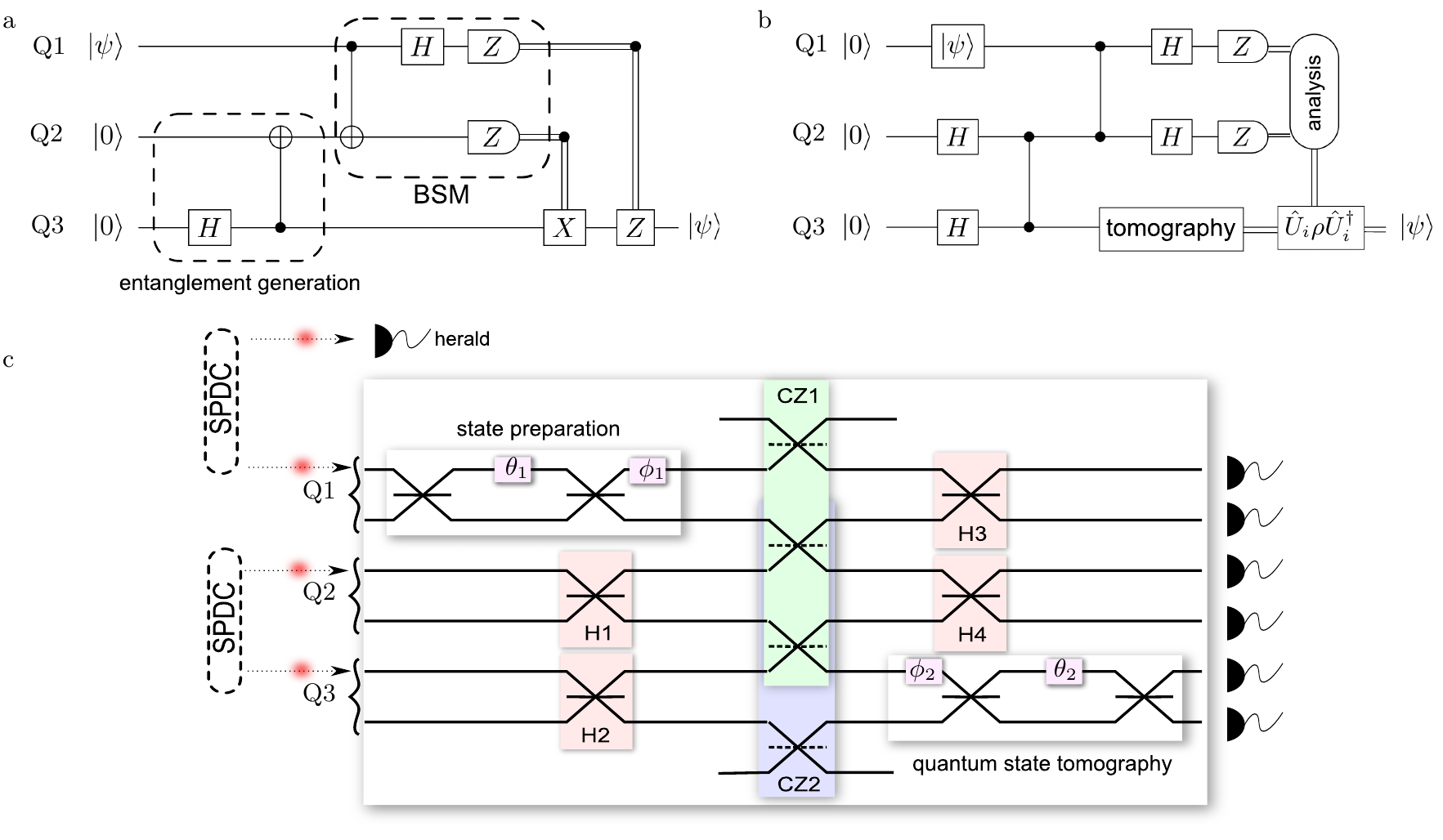}
\caption{{\bf Quantum teleportation and photonic chip realization} (a) Circuit diagram of a general quantum teleportation scheme. (b) In our experiment we replace the two \textsc{cnot}-gates with two \textsc{c-phase}-gates and additional local Hadamard operations. The results of the Bell state measurement are used in post-processing as part of quantum state tomography to recover the teleported qubit. (c) In the on-chip realization, three qubits are encoded using dual-rail logic in a silica-on-silicon integrated chip. Local Hadamard operations (\textsc{h1} to \textsc{h4}) are performed using  beam splitters of reflectivity 1/2 (solid lines) and the two cascaded \textsc{c-phase} gates (\textsc{cz1}, \textsc{cz2}) are implemented using four beam splitters of reflectivity 1/3 (dashed lines). State preparation and tomography are performed on chip using thermo-optically controlled phases $\theta$ and $\phi$.
  }
\label{fig:schematic}
\end{figure*}

Whilst photonic experiments were the first to realize quantum teleportation~\cite{Bouwmeester1997,Boschi1998}, demonstrations of this protocol in a waveguide architecture have been limited to fiber-based experiments~\cite{Nilsson2013,Marcikic2003}. Although there has been recent progress~\cite{martin2012}, no integrated photonic experiments have yet been able to demonstrate actual teleportation, due to the difficulty in realizing three photonic qubits on a sufficiently complex circuit~\cite{Metcalf2013}. In particular, integrated components require careful attention to fabricated deviations from design and the effects of increased and potentially unbalanced propagation loss. Experimental verification that integrated photonic circuits continue to perform well as their complexity increases is therefore of considerable interest.

In this letter, we demonstrate teleportation of a photonic qubit on an integrated waveguide device. We use a reconfigurable photonic chip to perform the teleportation protocol with the state encoding, entanglement preparation, Bell state analysis and state tomography all carried out on-chip.  We develop a theoretical model to account for all sources of possible error in the circuit and find good agreement with the measured teleported state fidelities, which exceed the average teleportation fidelity possible with a classical device. We identify the elements of this error budget relevant to scaling and find that improvements to chip characterisation and fabrication will be required to achieve high fidelity operation. For this work, the combination of high-heralding-efficiency single-photon sources, a low-loss silica waveguide architecture, as well as a careful element-wise characterization of the fabricated device enabled the successful operation of the first three-qubit logic operation on an integrated photonic platform.

The prototypical quantum teleportation circuit, shown in figure~\ref{fig:schematic}(a), aims to transfer the quantum state of an input qubit Q1 to the target qubit Q3. The protocol begins by generating a maximally entangled photonic resource, a Bell state, encoded on qubits Q2 and Q3. A two-qubit Bell state measurement (BSM) is then performed on input Q1 and one half of the entangled state (Q2), which does not reveal any information about the initial state of Q1. This measurement projects the target qubit Q3 onto the original input state via the non-local correlations of the entangled pair, modulo one out of four possible local unitary rotations. The specific unitary rotation needed to recover the initial state is uniformly distributed between the four possibilities, but can be identified by two classical bits describing the outcome of the BSM. In theory, this results in the perfect, unit-fidelity transfer of the initial unknown quantum state from Q1 to Q3 without any information being revealed about the state itself. Without the resource of entanglement or knowledge of the input state, the best achievable average fidelity is only 2/3~\cite{Massar1995}.

In our experiment, whilst the active feed-forward corrective rotation could in principle be implemented using the on-chip phase shifters $\phi_2$ and $\theta_2$, in reality these shifters do not have sufficient bandwidth to make this feasible. Instead, for the purposes of this demonstration, we use these phase shifters to perform quantum state tomography (QST) of the teleported qubit Q3 and replace the corrective step by a numerical rotation in post-processing. 

The waveguide circuit encodes three qubits using a dual-rail scheme as shown in figure~\ref{fig:schematic}(c). In this scheme, a single photon in the top rail represents the logical state $\ket{0_L}=\hat{a}^\dagger\ket{\textrm{vac}}=\ket{10}_{a,b}$ whereas one photon in the lower mode represents the logical state $\ket{1_L}=\hat{b}^\dagger\ket{\textrm{vac}}=\ket{01}_{a,b}$. Deterministic single-qubit operations are realized using linear beam splitters and thermo-optic phase shifters (see Methods). These elements allow us to implement reconfigurable single qubit unitary rotations, $\hat{U}(\theta,\phi)=e^{-i\phi\hat{\sigma}_z/2}e^{-i\theta\hat{\sigma}_y/2}$, which we use in the state preparation and state tomography stages depicted in figure~\ref{fig:schematic}(c) (see Methods).

The concatenated \textsc{c-phase}-gates at the heart of the circuit are implemented using the probabilistic scheme proposed by Ralph~\cite{Ralph2004}. In this scheme, effective two-photon interactions are induced via post-selection of one photon being detected in each of the three qubit modes. This occurs with probability 1/27, which sets the overall success probability of the circuit. 

\begin{figure*}[t]
\centering
\includegraphics[width=0.85\textwidth]{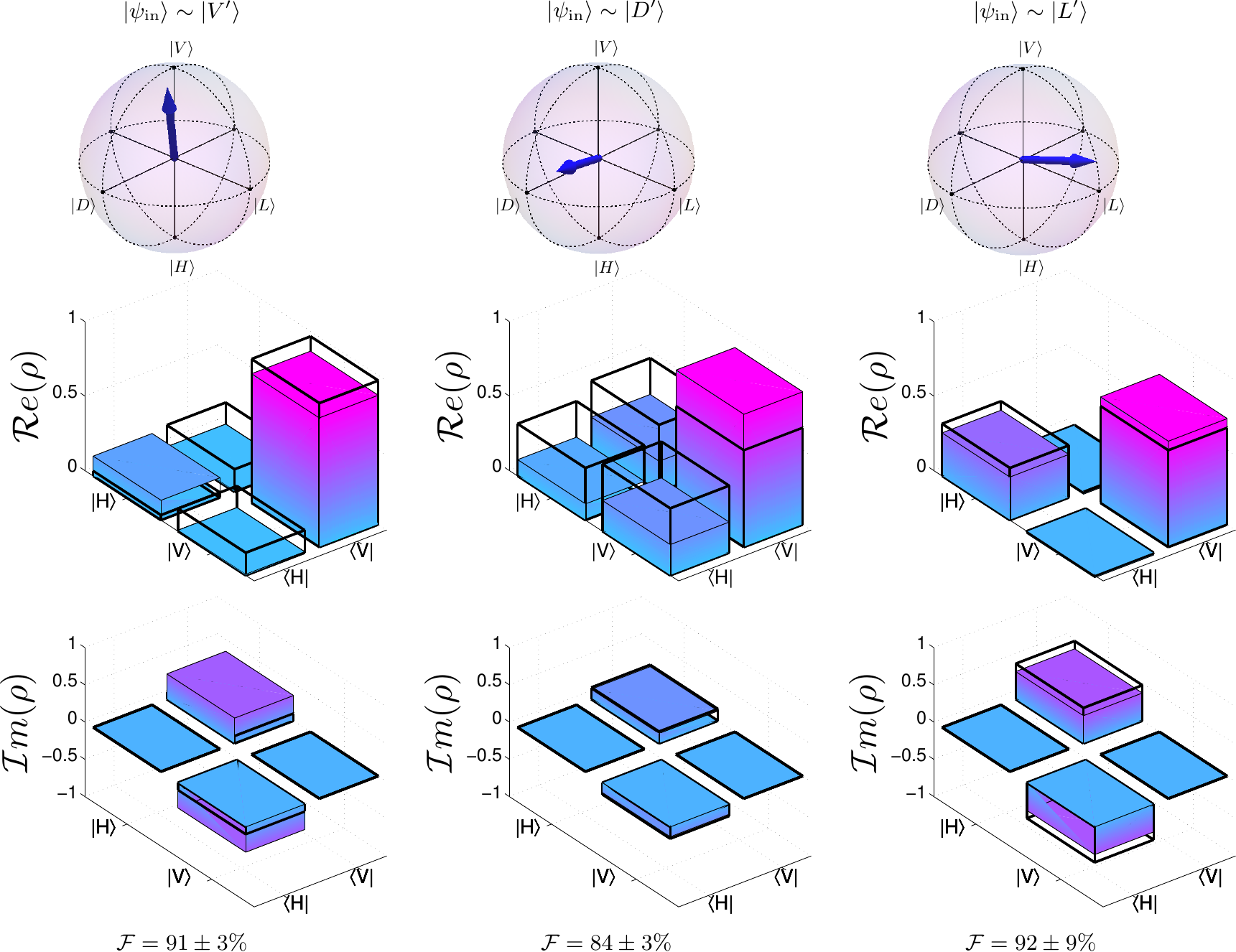}
 \caption{{\bf Reconstructed density matrices of the teleported states} The initial qubit states on Q1 for each of three trials (columns) are depicted on the bloch sphere (top) and as real and imaginary parts of a density matrix (black wire frames, middle). The final teleported states on Q3 are reconstructed using on-chip quantum state tomography and then transformed by optimal state-independent rotations in post-processing (coloured bars). The fidelity, $\mathcal{F}$, between the initial and final state shown is calculated (bottom). Representative data here are for experiments with a $\ket{\Psi^+}$ Bell state measurement outcome. Similar reconstructed states for all four Bell state measurement outcomes are found in the Supplementary Information.}
  \label{fig:data}
\end{figure*}

We now describe the main experiment showing on-chip quantum teleportation. Three single photons generated by two parametric down-conversion sources are coupled into the UV-written, silica-on-silicon photonic chip (see Methods). An array of six avalanche photodiodes (APDs) monitor all output modes. We identify successful teleportation events as runs in which one photon is detected for each of the three qubits, as well as in an ancillary heralding arm. Four-fold detection coincidences are registered using an FPGA. Data is collected for three linearly independent input states, $\ket{\psi_\textrm{in}}=\hat{U}(\theta_1,\phi_1)\ket{0_L}$, coded on Q1 by appropriately setting $\theta_1$ and $\phi_1$. For each input state, $\theta_2$ and $\phi_2$ are adjusted to perform  projective measurements of Q3 in three different bases, allowing us to reconstruct the teleported state via maximum-likelihood tomography.

Each input state is teleported to one of four possible output states depending on the outcome of the BSM: $\ket{\psi_\textrm{out}}=U_i\ket{\psi_\textrm{in}}$, where the rotation $U_i$ is uniquely determined by the BSM outcome. For an ideal circuit, $U_i$ corresponds to an element of $\{\hat{\sigma}_z,\mathbb{1},i\hat{\sigma}_y,\hat{\sigma}_x\}$. A realistic circuit deviates from the ideal case due to fabrication imperfections. In general, the rotations that optimize the teleportation fidelity must be found by characterizing the actual device. To do so, we use a numerical model to simulate the teleported output state at the point immediately prior to the QST stage on Q3. The simulation uses ideal indistinguishable Fock-state inputs and critically relies on being able to perform an element-wise characterisation of our circuit in order to remove the effects of the on-chip state encoding and tomography stages from this analysis. We then numerically find the unitary rotations, $U_i'$, which maximize the average output state fidelity over 10,000 randomly chosen input states. We note that the resulting rotations are found independently of the primary experimental data and depend only on the classical circuit characterisation. In order to verify successful teleportation, we apply these rotations to the reconstructed state of Q3, $\hat{\rho}_\textrm{out}$, in post-processing and calculate the fidelity to the input states, $\mathcal{F}=\bra{\psi_\textrm{in}}\hat{U}'_i\hat{\rho}_\textrm{out}\hat{U}_i'^{\dagger}\ket{\psi_\textrm{in}}$.   

\begin{figure*}
  \centering
\includegraphics[width=0.9\textwidth]{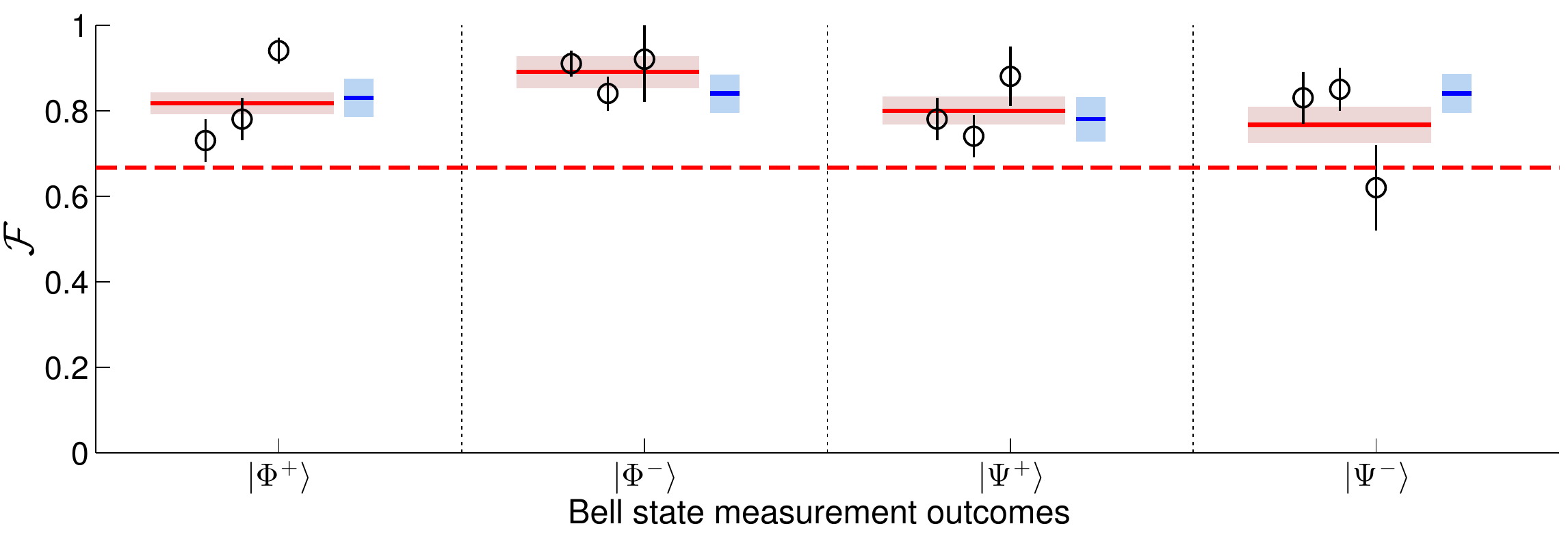}
  \caption{{\bf Measured and simulated fidelity of on-chip quantum teleportation} The measured fidelity of the three teleported states ($|V'\rangle$, $|D'\rangle$ and $|L'\rangle$ from left to right respectively) for each outcome of the BSM are plotted with circles. Errors are calculated using a Monte-Carlo method taking account of Poissonian counting statistics and uncertainties in the characterized measurement operators and input states. The red shaded areas show the measured fidelity averaged over all three input states and clearly exceeds the best average classical fidelity given by the red dotted line. The blue shaded areas show the predicted average fidelity taking account of imperfect circuit fabrication, higher order photon emission and residual photon distinguishability. The error on these predictions are estimated using a Monte-Carlo method over a range of different input states and model parameters.}
\label{fig:fidPlot}
\end{figure*}

The teleportation results for the BSM outcome $\ket{01}$ are summarized in figure~\ref{fig:data}. The experiment was performed using three linearly independent input states close to the three orthogonal axes of the Bloch sphere (see Supplementary Information for details). We label them $\ket{V'}$, $\ket{D'}$ and $\ket{L'}$, according to the closest lying axis state (figure~\ref{fig:data}(a-c)). Four-fold coincidences were registered in a specific BSM outcome at approximately 5 mHz and around 100 coincidence counts were collected for each measurement setting and BSM outcome. The initial input states are recovered with an average fidelity of $89\pm3\%$ which exceeds the classical limit of 2/3 by more than 6 standard deviations. Since any state on the Bloch sphere can be generated using only linear combinations of these three states, these results are sufficient to conclude that this chip is capable of teleporting general quantum states with high fidelity. The results from all four BSM outcomes are shown in figure~\ref{fig:fidPlot}.  In each case, the average fidelity for the three input states (shaded red boxes) is higher than the classical limit (dashed red line), irrespective of which classical outcome is obtained. Unbalanced propagation loss in combination with beam splitter ratios which deviate from their designed reflectivity mean that the four Bell state outcomes occur with non-equal probability resulting in variations in the number of coincidence events recorded for each measurement.

The beam splitter reflectivities and behaviour of the phase shifters were characterized to determine the unitary rotations and to correctly simulate the performance of the experiment. Recent proposals to characterize the behaviour of linear optical circuits have treated the devices as a `black box', returning the overall transfer matrix of the network without reference to any information about the geometry of the underlying circuit~\cite{Rahimi-Keshari2013,Laing2012a}. For experiments using a reprogrammable circuit, however, it is useful to be able to characterize the individual linear optical elements so that it is not necessary to perform a full characterization for every configuration of the circuit. A full loss-tolerant characterization of the twelve beam splitters was performed making use of transversely scattered light from the waveguide~\cite{Metcalf2013}. The voltage-controlled phase-shifters were calibrated in pairs to account for thermal cross-talk between heaters in close proximity (see Methods).

The sources responsible for the reduction in fidelity were investigated using a full theoretical model taking into account the experimentally characterized beam splitter ratios and interferometer phases, higher-order photon emission, photon distinguishability and propagation losses. This model was used to calculate the expected click statistics on Q3 for given BSM outcomes and phase-shifter settings. These simulated clicks were substituted into our data analysis routine, including the appropriate unitary rotation, to predict the expected teleportation fidelity for different input states. This analysis reveals a slight input-state dependent fidelity caused by the deviation of the fabricated beam splitting ratios away from their ideal values (see Supplementary Information). The predicted teleportation fidelity averaged over all input states are summarized in figure~\ref{fig:fidPlot} as shaded blue boxes. 

The model allows us to quantify the reduction in achievable fidelity which can be attributed to different parts of the experiment. The non-ideal beam-splitting ratios reduce the quality of the entanglement in the circuit and immediately reduce the achievable fidelity to around 90\% - a figure which still exceeds the cloning limit of $\mathcal{F}=5/6$ required for secure communication~\cite{Buzek1996,Grosshans2001} but is still below the fidelity of $99\%$ thought to be required for a fault-tolerant quantum computer~\cite{Kok2007}.

The non-ideal nature of the photon sources is found to be responsible for the remaining observed reduction in measured fidelity, since the protocol relies critically on high-visibility quantum interference between all three photons. We estimate the magnitude of the residual photon distinguishability and higher-order photon emission from our squeezed sources and add these to the theoretical model (see Methods).

This error analysis highlights three key impediments to future scaling: photon sources, imperfect photonic chips and loss. Whilst the success of this experiment relies on our development of high-quality single photon sources with exceptional heralded state purity and heralding efficiency, the absence of a true on-demand single photon source continues to limit the achievable fidelity. This ubiquitous problem in quantum optics experiments is being addressed by the recent development of low-loss waveguided sources~\cite{Spring2013,Eckstein2011}. Secondly, moving to larger on-chip experiments will place stringent demands on photonic circuit performance. Fabricated beam splitters and phase shifters will inevitably show some deviation from their designed parameters. Recent progress demonstrating tighter control of fabricated optical components is promising~\cite{Crespi2013}. However, as more components are integrated onto circuits, robust characterisation methods together with active circuit control~\cite{smith2009,Shadbolt2011} will be required to identify, and then correct, remaining deviations from design. Finally, scattering losses not only reduce data rates but exacerbate the issues identified above, already creating challenges in small-scale experiments~\cite{Spring2012}. Lower loss dielectrics, in addition to loss-tolerant protocols, will aid future work.  

In conclusion, we present the on-chip implementation of a three-qubit quantum circuit, successfully teleporting three linearly independent quantum states.  We have modelled sources of error and their effect on the teleportation fidelity.  Whilst integrated photonics offers a feasible route to larger-scale LOQC, this work shows that continuing advances in waveguide and photon source technology will be critical in addressing the challenges posed by complex integrated circuits - coping with the increased loss and identifying and correcting component errors.

\section*{Methods}
\subsection*{Device fabrication}
\footnotesize{
The waveguide circuit used in this work was fabricated by the direct UV-writing technique utilising silica slab waveguides deposited on a silicon substrate~\cite{smith2009}. The individual waveguides were written by focusing a continuous-wave UV laser (244 nm wavelength) onto the germanium doped silica photosensitive waveguide core and translating the laser beam transversely to the surface normal with computer-controlled 2D motion. Waveguides of typical 4.5$\mu$m$\times$4.5$\mu$m dimension were formed as a result of UV-induced permanent refractive index change inside the photosensitive waveguide core layer. The UV-writing process enables creation of complex networks utilising compact X couplers~\cite{kundys2009}, the splitting ratio of which can be selected by adjusting the waveguide crossing angle during the UV-writing process. Compact X-couplers have a number of advantages over more traditional directional couplers i.e. compact footprint, low guiding loss and more stable coupling ratio~\cite{kundys2009}. The thermo-optic phase shifters utilize a small NiCr electrode (0.35\,$\mu$m $\times$ 50\,$\mu$m $\times$ 2.5\,mm, 0.85\,kOhm electrical resistance) deposited directly over one of the waveguides through which a current can be passed. The passive stability of the interferometers with the phase-shifters set to a constant voltage was measured to be less than $1^{\circ}$ over 24 hours and achieved a repeatability error of less than $5^\circ$ when the voltage settings were changed.}

\subsection*{Single photon source}
\footnotesize{
An 80\,MHz Ti:Sapphire oscillator (Mai-Tai, Spectra Physics) produces 100\,fs pulses at 830\,nm which are upconverted to 415\,nm in a 700$\,\mu\textrm{m}$  $\beta-\textrm{BaB}_2\textrm{O}_4$ (BBO) crystal cut for type-I second-harmonic generation. This is split on a 50:50 beam splitter and used to pump two 8\,mm-long AR-coated Potassium Dihydrogen Phosphate (KDP) crystals phase-matched for degenerate type-II collinear parametric down-conversion. The source is spectrally factorable, improving the heralding efficiency achieved when interference filters (Semrock, $\Delta\lambda=3$\,nm) are used to match the bandwidths of the broad and narrowband daughter photons. With the filters in place we achieve a four photon coincidence rate of 20\,Hz and two-photon fidelities of 0.98 (narrowband-narrowband) and 0.97 (narrowband-broadband), as measured using a Hong-Ou-Mandel interferometer (see Supplementary Information).}

\subsection*{Circuit characterisation}
The on-chip state preparation and tomography elements each consist of two thermo-optic phase shifters embedded in a Mach-Zehnder Interferometer. Each pair of phase shifters are situated close enough together that residual heat from one can slightly affect the other. The effect of this cross-talk must be characterised to accurately determine the behaviour of the circuit at different phase settings. Bright light was alternately coupled into the two input modes of Q1 whilst the power on the two output modes was monitored using photodiodes. The voltages applied to $\theta_1$ and $\phi_1$ are both varied to map out the 2D response of the shifters on the effective phase change within their respective interferometers (see Supplementary Information).  A similar 2D response is measured for the two phase shifters on Q3 and the zero-phase offset of the central interferometer on Q2 is also calculated. We find a small amount of cross-talk at the upper ranges of applied heater power and account for this in our analysis.

\subsection*{Programmable unitary rotations}
Generating and measuring arbitrary dual-rail qubit states, requires control over only two parameters - a phase shift and a tunable beam splitter -  $\hat{U}_\textrm{prepare}(\phi_1,\theta_1)=e^{-i\phi_1\hat{\sigma}_z}e^{-i\theta_1\hat{\sigma}_y/2}$ and $\hat{U}_\textrm{measure}(\phi_2,\theta_2)=e^{-i\theta_2\hat{\sigma}_y}e^{-i\phi_2\hat{\sigma}_z/2}$ - where $\hat{\sigma}_{x,y,z}$ are the Pauli matrices. A tunable beam splitter may be realized by embedding a phase shifter within a Mach-Zehnder Interferometer (see Supplementary Information). In the ideal case, the MZI is composed of two $\eta=1/2$ beam splitters. As the beam splitter reflectivity deviates away from this ideal, we no longer implement a perfect tunable beam splitter, restricting the range of accessible input states and measurement operators (see Supplementary Information). The limited tuning range of our thermal phase shifters ($0<\{\theta,\phi\}\lesssim1.6$) combined with our non-ideal beam splitters restrict us to preparing and measuring quantum states over only one octant of the Bloch Sphere. 

\subsection*{Simulating teleportation fidelities}
The expected click statistics of the photonic circuit were simulated by propogating input Fock states through a linear transfer matrix generated using the measured beam splitter ratios and interferometer phase offsets. For a given input of $N$ photons distributed across the three input modes, the output probability distribution of clicks was obtained by calculating the permanents of specific $N\times N$ sub-matrices of the transfer matrix (see Supplementary Material). However, the photon sources used in this work generate two-mode squeezed states: $\ket{\Psi_{\textrm{PDC}}}=\sqrt{1-\lambda^2}\sum_{n=0}^{\infty}\lambda^n\ket{nn}$. We thus model the input state as a mixture of the ideal three-photon Fock state input ($\ket{1,1,1}$) and the first set of higer order terms ($\ket{1,2,2}, \ket{2,1,1}$), weighted by the squeezing parameter of the photon sources measured via the conditional second-order correlation function ($\lambda^2\sim0.03$). The output probability distribution when one photon is distinguishable from the others is given by the incoherent sum of a single and an $N-1$ photon input (see Supplementary Information). These output photon distributions are weighted by the distinguishability of our photon source through the measured reduction in expected Hong-Ou-Mandel dip visibility as discussed in the single photon source Methods section.

\section*{Acknowlegements}
We thank S.~Tanzilli for insightful comments on the manuscript. This work was supported by the EPSRC(EP/K034480/1, EP/H03031X/1, EP/H000178/1), the EC project SIQS, the Royal Society, and the AFOSR EOARD. XMJ, WSK and NKL are supported by EC Marie Curie fellowships (PIIF-GA-2011-300820, PIEF-GA-2012-331859).
\section*{Author Contributions}
\footnotesize{
B.J.M.\, J.B.S.\, P.C.H.\, N.T.-P.\, N.K.L.\ and I.A.W.\ all contributed to designing and setting up the experiment.  B.J.M.\ performed the experiment.  J.B.S.\ designed the FPGA electronics and helped with data taking.  D.K.\ and J.C.G.\ fabricated the waveguide device.  X.-M.J., W.S.K., M.B., P.C.H.\, J.B.S., B.J.M.\ all contributed to the analysis of the data. B.J.M wrote the manuscript with input from all authors. B.J.S, P.G.R.S\ and I.A.W.\ conceived the work and supervised the project.   
}

\end{document}